\documentclass[showkeys,superscriptaddress,floatfix,aps,10pt,prd]{revtex4-2}
\usepackage{graphicx,epstopdf}
\usepackage[caption=false]{subfig}
\usepackage{appendix}
\usepackage[T1]{fontenc}
\usepackage{lmodern}
\usepackage[dvipsnames,x11names]{xcolor}
\usepackage[colorlinks=true,linkcolor=ForestGreen,citecolor=ForestGreen,urlcolor=NavyBlue]{hyperref}
\usepackage[sort&compress]{natbib}
\usepackage{morefloats}
\usepackage[pdf]{pstricks}
\usepackage{amsmath}
\usepackage{amssymb}
\usepackage{amsfonts}
\usepackage{rotating}
\usepackage{cancel}
\usepackage{mathtools}
\usepackage{bbm}
\usepackage{dsfont}
\usepackage{bbold}
\usepackage{multirow}
\usepackage{ulem}
\usepackage{physics}
\usepackage{orcidlink}
\usepackage{colortbl}

\def\xslash{x\!\!\!\slash }

\begin{document}

\title{Electromagnetic properties of the $D_{s1}^{+}(2460)$, $D_{s1}^{+}(2536)$, and their bottom partners in a molecular configuration}

\author{Ula\c{s}~\"{O}zdem\orcidlink{0000-0002-1907-2894}}%
\email[]{ulasozdem@aydin.edu.tr }
\affiliation{ Health Services Vocational School of Higher Education, Istanbul Aydin University, Sefakoy-Kucukcekmece, 34295 Istanbul, 
T\"{u}rkiye}

 
\begin{abstract}
We investigate the electromagnetic properties of the axial-vector molecular states $D^* K$, $DK^*$, $B^* K$, and $BK^*$, which are used to model the charmed states $D_{s1}^{+}(2460)$, $D_{s1}^{+}(2536)$, and their bottom partners with quantum numbers $J^P = 1^+$. To our knowledge, this presents the first comprehensive calculation of the magnetic and quadrupole moments for these specific molecular configurations. Employing the QCD light-cone sum rule method with molecular-type interpolating currents, we compute these moments and perform a detailed flavor decomposition to reveal the internal distribution of the electromagnetic charge and spin. Our results demonstrate that the light up and down quarks dominate the electromagnetic response, with negligible contributions from the heavy quarks. The $D^* K$ and $B^* K$ states exhibit negative quadrupole moments and slightly oblate charge distributions, whereas the $DK^*$ and $BK^*$ states possess positive quadrupole moments and prolate distributions, with significant contributions from the strange quark. The predicted moments provide benchmarks for lattice QCD calculations and are testable through their influence on radiative transitions and photo- and electro-production observables at high-luminosity facilities, offering crucial insights into the internal structure and nature of these axial-vector states.
\end{abstract}

\maketitle

\section{motivation}\label{motivation}

The $D_{s_0}(2317)$ state was first observed by the BaBar Collaboration~\cite{BaBar:2003oey} in the $D_s^+ \pi^0$ invariant mass spectrum produced in $e^+e^-$ annihilation in 2003. In the same year, the $D_{s1}(2460)$, considered the spin partner of $D_{s_0}(2317)$ within the framework of heavy quark spin symmetry, was reported by the CLEO Collaboration~\cite{CLEO:2003ggt}. These observations were subsequently confirmed by several other experiments~\cite{Belle:2003guh, BaBar:2004yux,BaBar:2006eep}. In addition, the $D_{s1}(2536)$ charm-strange state was measured earlier by the Belle Collaboration~\cite{ARGUS:1989zue}. Following the discovery of these states, extensive experimental studies have been performed~\cite{Belle:2003kup,BaBar:2004gtd, BaBar:2003cdx, Akhmedov:2005np}. The charm-strange mesons $D_{s_0}(2317)$, $D_{s1}(2460)$, and $D_{s1}(2536)$ have been measured with the following properties. The $D_{s_0}(2317)$, with quantum numbers $J^P = 0^+$, has a mass of $2317.8 \pm 0.5$~MeV and a width constrained to be less than $3.8$~MeV. The $D_{s1}(2460)$, assigned $J^P = 1^+$, exhibits a mass of $2459.5 \pm 0.6$~MeV with a width smaller than $3.5$~MeV. Finally, the $D_{s1}(2536)$, also with $J^P = 1^+$, has a precisely measured mass of $2535.11 \pm 0.06$~MeV and a narrow decay width of $0.92 \pm 0.05$~MeV.   
Simultaneously, numerous theoretical studies have focused on the internal structure of these charm-strange states. Notably, the masses of $D_{s_0}(2317)$ and $D_{s1}(2460)$ are approximately 100~MeV lower than the predictions of conventional quark models. This significant discrepancy has also been observed in lattice QCD simulations, posing challenges for both theorists and experimentalists. This phenomenon is commonly referred to as the “low-mass problem,” and similar deviations have been reported for states such as $\Lambda_c(2940)$ and $X(3872)$.  The low-mass problem is important because it highlights a significant deviation from the expectations of the conventional quark model. This discrepancy points to the limitations of traditional hadron classifications and provides one of the strongest indications for the existence of exotic, non-conventional structures such as hadronic molecules or tetraquarks.

To address this issue and gain insight into the structure of these mesons, a variety of theoretical interpretations have been proposed. For instance, $D_{s_0}(2317)$ and $D_{s1}(2460)$ have been studied as compact tetraquark configurations \cite{Nielsen:2005ia, Wang:2006uba, Chen:2004dy, Dmitrasinovic:2005gc, Cheng:2003kg, Terasaki:2003qa, Dmitrasinovic:2004cu, Kim:2005gt, Zhang:2018mnm}, as molecular states\cite{vanBeveren:2003kd, Liu:2012zya, Barnes:2003dj, Roca:2025lij, Kong:2021ohg, Kolomeitsev:2003ac, Szczepaniak:2003vy, Hofmann:2003je, Gamermann:2006nm, Flynn:2007ki, Faessler:2007gv, Guo:2009ct, Xie:2010zza, Cleven:2010aw, Wu:2011yb, Guo:2015dha, Albaladejo:2016hae, Du:2017ttu, Guo:2018tjx, Albaladejo:2018mhb, Wu:2019vsy, Gregory:2021rgy, Wang:2012bu, Huang:2021fdt,Chen:2016spr,Liu:2019zoy, Zhou:2025yjb, Zhou:2025rpb},  as conventional quark-antiquark mesons \cite{Bardeen:2003kt, Deandrea:2003gb, Dai:2003yg, Sadzikowski:2003jy, Cahn:2003cw, Hwang:2004cd, Simonov:2004ar, Cheng:2014bca, Song:2015nia, Cheng:2017oqh, Luo:2021dvj, Zhou:2020moj, Alhakami:2016zqx}, or as mixed states containing both two- and four-quark components \cite{Vijande:2006hj, Maiani:2004vq, Dai:2006uz, Browder:2003fk}. Additionally, alternative works suggest that these states could be understood as molecular bound states formed by two mesons. Regarding the $D_{s1}(2536)$, several studies treat it as a $DK^\ast  $ molecular tetraquark, arising naturally from coupled-channel interactions within the molecular framework \cite{Lin:2024hys,Gamermann:2007fi,Guo:2006fu,Guo:2006rp}.  Alongside open-charm molecular states, bottom-strange systems with quantum numbers $I(J^P) = 0(0^+)$ and $0(1^+)$ have also attracted considerable attention\cite{Guo:2006fu, Guo:2006rp, Faessler:2008vc}. Although experimental data are currently limited, theoretical studies indicate the possible existence of bound states such as $BK$ and $B^\ast  K$ with energies lying below their respective thresholds. More recently, the LHCb collaboration reported the observation of two structures that may be identified as $B_{sJ}(6064)$ and $B_{sJ}(6114)$ if they decay directly into $B^+ K^-$ \cite{LHCb:2020pet}. Alternatively, these states could correspond to $B_{sJ}(6109)$ and $B_{sJ}(6158)$ when decaying through the $B^{\ast   \pm} K^\mp$ channel. Notably, the measured mass of $B_{sJ}(6158)$ is $6158 \pm 4 \pm 5$ MeV, which lies close to the $BK^\ast  $ threshold. Therefore, the study of the molecular states $BK$, $B^\ast  K$, and $BK^\ast  $ constitutes a timely and relevant research topic.

In other words, despite the numerous investigations into the internal structures of $D_{s1}(2460)$, $D_{s1}(2536)$, and their bottom-sector counterparts using various theoretical approaches, the resulting conclusions have not been fully consistent. Therefore, it remains important to further study these open-charm and bottom states to clarify their structural and electromagnetic properties. Despite growing phenomenological support for such interpretations, the experimental distinction between compact tetraquark states and extended molecular configurations remains a formidable challenge.  In our previous work~\cite{Ozdem:2025hmb}, we investigated hidden-charm vector molecular states, such as $D\bar{D}_1(2420)$ and $D^*\bar{D}^*(2400)$, using the QCD light-cone  sum rule framework. However, in the present study we focus on open-flavor axial molecules, $D_{s1}^{+}(2460)$, $D_{s1}^{+}(2536)$, and their bottom partners. These systems are flavor-asymmetric even in the chiral limit, possess different quantum numbers ($J^P=1^+$), and require distinct interpolating currents. Although the methodological steps are similar, the physical systems, quark-flavor structure, and phenomenological implications are entirely different. Hence, the present study constitutes an independent investigation and addresses questions not covered in our previous hidden-charm analysis. 
To this end, electromagnetic properties—particularly magnetic dipole moments—emerge as valuable diagnostics. These observables are directly sensitive to the spatial distribution of charges and spins inside hadrons and provide nontrivial insights into the underlying quark-gluon dynamics. In processes such as photoproduction or radiative transitions, magnetic moments influence the cross-sections and angular distributions, making them testable in high-luminosity experiments. In this work, we employ the QCD light-cone sum rules (LCSR) approach~\cite{Chernyak:1990ag, Braun:1988qv, Balitsky:1989ry} to compute the magnetic and quadrupole  moments of the $D^\ast K$, $DK^\ast$, $B^\ast K$, and $BK^\ast$ molecular states, with quantum numbers $J^{P} = 1^{+}$. LCSR offers a unique advantage by incorporating non-perturbative effects via photon distribution amplitudes, while maintaining the perturbative treatment of short-distance quark-gluon interactions. By correlating hadronic and QCD representations of the relevant correlation functions under the assumption of quark-hadron duality, we derive sum rules for the magnetic moments of these exotic molecular configurations. Our results aim to provide a deeper understanding of their internal structure and contribute to the growing body of evidence that supports the molecular interpretation of certain open heavy-flavor states. 
Several works in the literature have investigated the electromagnetic properties of singly-heavy hadrons, particularly those whose internal structures and characteristics remain under active debate~\cite{Ozdem:2024ydl,Ozdem:2024pyb,Ozdem:2023edw,Ozdem:2023okg,Ozdem:2023eyz,Ozdem:2022ydv,Ozdem:2021vry,Azizi:2021aib,Azizi:2018jky,Azizi:2018mte}.

This paper is organized as follows. Section~\ref{formalism} establishes the theoretical framework based on QCD light-cone sum rules, deriving the correlation functions consistently in both hadronic and quark-gluon representations and providing analytical expressions for the magnetic and quadrupole moments. Section~\ref{numerical} presents a detailed numerical analysis of these electromagnetic moments, along with an interpretation of the results. Finally, Section~\ref{sum} summarizes the main conclusions and highlights the significance of the findings.

 \begin{widetext}
 
\section{Theoretical Foundations}\label{formalism}

We present in this section the derivation of LCSR to compute the magnetic and quadrupole moments of the $D^\ast  K$, $DK^\ast  $, $B^\ast  K$, and $BK^\ast  $ molecular states (hereafter $T_Q$). The procedure starts by considering the following correlation function, which constitutes an essential component of the method:
\begin{equation}
 \label{edmn01}
\Pi _{\alpha \beta }(p,q)=i\int d^{4}x\,e^{ip\cdot x}\langle 0|\mathcal{T}\{J_{\alpha}(x)
J_{\beta }^{ \dagger }(0)\}|0\rangle_{F}.  
\end{equation}
The symbol $F$ indicates the external background electromagnetic field, while $J_{\alpha}(x)$ refers to the interpolating current of the $T_Q$ states. The interpolating currents for these molecular tetraquark states are formulated as \cite{Zhou:2025yjb}:
\begin{align}
\label{curr1}
J_{\alpha }^{1}(x) &= \frac{1}{\sqrt{2}} \Big\{\big[\bar u^{a} (x) \gamma_\mu Q^a (x)\big]\big[\bar s^b (x)  i \gamma_5  u^{b} (x) \big] +
\big[\bar d^{a} (x) \gamma_\mu Q^a (x)\big]\big[\bar s^b (x)  i \gamma_5  d^{b} (x) \big] \Big\},\\
J_{\alpha }^{2}(x) &= \frac{1}{\sqrt{2}} \Big\{\big[\bar u^{a} (x) i\gamma_5 Q^a (x)\big]\big[\bar s^b (x)   \gamma_\mu  u^{b} (x) \big] +
\big[\bar d^{a} (x) i\gamma_5 Q^a (x)\big]\big[\bar s^b (x)  \gamma_\mu  d^{b} (x) \big] \Big\},
\label{curr2}
\end{align}
where $a$ and $b$ are color indices, $Q$  stands for $c$ or $b$ quark
fields. 
The current $J_{\alpha}^{1}(x)$ is designed to create the state dominantly through $[\bar{q}\gamma_{\mu}Q] \otimes [\bar{s}i\gamma_{5}q]$ configurations. This combination effectively carries the quantum numbers of a vector heavy-light meson ($D^{*}/B^{*}$) and a pseudoscalar strange meson ($K$), coupling to total spin-parity $J^{P} = 1^{+}$. This makes it the natural choice for probing the $D^{*}K$ and $B^{*}K$ molecular candidates.
Conversely, the current $J_{\alpha}^{2}(x)$, of the form $[\bar{q}i\gamma_{5}Q] \otimes [\bar{s}\gamma_{\mu}q]$, is tailored to overlap primarily with $[\bar{q}i\gamma_{5}Q] \otimes [\bar{s}\gamma_{\mu}q]$ structures. This corresponds to a pseudoscalar heavy-light meson ($D/B$) and a vector strange meson ($K^{*}$), also yielding $J^{P} = 1^{+}$. It is therefore the appropriate operator for the $DK^{*}$ and $BK^{*}$ molecular channels.
The choice of these specific color-singlet bilinear structures is crucial. It explicitly assumes that the dominant Fock component of the physical state is a loosely bound hadronic molecule, where the constituent mesons largely retain their individual identities. This is in stark contrast to interpolating currents built, for instance, from diquark-antidiquark pairs ($[q^T C Q] [\bar{s} C \bar{q}^T]$), which would be more suitable for modeling compact tetraquark cores. While both molecular- and tetraquark-type currents can, in principle, couple to the same physical state, the molecular currents used here are expected to be more efficient for states that lie very close to the corresponding meson-meson thresholds—a key characteristic of the states under investigation~\cite{Zhou:2025yjb}. This operator choice allows the LCSR method to most effectively capture the electromagnetic properties arising from the extended spatial structure and the coherent sum of the constituent meson moments expected in a molecular picture.

The interpolating currents in Eqs. (\ref{curr1}) and (\ref{curr2}) allow us to examine the electromagnetic features of the molecular states using LCSR. The approach is summarized in the following procedure:

\begin{itemize}
\item The first stage entails representing the correlation function in terms of hadronic quantities, such as the mass, magnetic moment, and form factors. This approach is commonly known as the ``hadronic formulation.''
    
\item In the next step, the correlation function is represented using quantities associated with quark and gluon dynamics, together with the relevant distribution amplitudes. This approach is commonly termed the ``QCD formulation.''

\item Ultimately, the two representations are reconciled based on quark-hadron duality. Unwanted effects are suppressed through a double Borel transformation and continuum subtraction, leading to the derivation of the sum rules for the relevant physical parameters.
\end{itemize}

In accordance with the procedure described above, our analysis starts by examining the hadronic formulation of the relevant molecular states. 
The general procedure closely follows the framework developed in Ref. \cite{Ozdem:2025hmb}, where similar correlation functions were analyzed for different flavor-symmetric tetraquark configurations. It is outlined here for completeness and to maintain self-consistency, as the present investigation deals with asymmetric heavy–light configurations that require distinct interpolating currents, quark-flavor arrangements, and slightly modified normalization conventions.

\subsection{Hadronic formulation}

The hadronic formulation of the correlation function is obtained by introducing a complete set of intermediate states matching the quantum numbers of the $T_Q$ states associated with the interpolating currents. 
At this stage, after the necessary mathematical manipulations, the correlation function can be expressed as:
\begin{align}
\label{edmn04}
\Pi_{\alpha\beta}^{Had}(p,q) &= \frac{\langle 0 \mid J_\alpha(x) \mid T_Q(p, \varepsilon^i) \rangle}{p^2 - m_{T_Q}^2} \langle T_Q(p, \varepsilon^i) \mid T_Q(p+q, \varepsilon^f) \rangle_F \frac{\langle T_Q(p+q, \varepsilon^f) \mid J_\beta^\dagger(0) \mid 0 \rangle}{(p+q)^2 - m_{T_Q}^2} \nonumber\\
&+ \mbox{continuum and higher states}.
\end{align}

The preceding expression highlights the necessity of providing explicit forms for the matrix elements involved in the correlation function. These elements, representing the overlap between the relevant hadronic states and the external electromagnetic field, are detailed below \cite{Brodsky:1992px}:
\begin{align}
\label{edmn05}
\langle 0 \mid J_\alpha (x) \mid T_Q (p, \varepsilon^i) \rangle &=  \lambda_{T_Q} \varepsilon_\alpha^i\,,\\
\langle T_Q (p+q, \varepsilon^{f}) \mid J_{\beta }^{\dagger } (0) \mid 0 \rangle &= \lambda_{T_Q} \varepsilon_\beta^{* f}\,,\\
\langle T_Q(p,\varepsilon^i) \mid  T_Q (p+q,\varepsilon^{f})\rangle_F &= - \varepsilon^\gamma (\varepsilon^{i})^\mu (\varepsilon^{f})^\nu
\Big[ G_1(Q^2)~ (2p+q)_\gamma ~g_{\mu\nu}  
+ G_2(Q^2)~ ( g_{\gamma\nu}~ q_\mu -  g_{\gamma\mu}~ q_\nu)
\nonumber\\ 
&
- \frac{1}{2 m_{T_Q}^2} G_3(Q^2)~ (2p+q)_\gamma
q_\mu q_\nu  \Big]. \label{edmn06}
\end{align}
In this expression, 
$\lambda_{T_Q}$ stands for the current coupling, 
$\varepsilon_\alpha^{\mathrm{i}}$ ($\varepsilon_\beta^{*\mathrm{f}}$) denotes the polarization vector of the initial (final) $T_Q$ tetraquark, 
$\varepsilon^\gamma$ is the photon polarization vector, and 
$G_i(Q^2)$ correspond to the electromagnetic form factors, where 
$Q^2 = -q^2$ represents the squared momentum transfer.

By employing the previously defined expressions, the correlation function in its hadronic formulation is expressed as:
\begin{align}
\label{edmn09}
 \Pi_{\alpha\beta}^{Had}(p,q) &=  \frac{\varepsilon_\rho \, \lambda_{T_Q}^2}{ [m_{T_Q}^2 - (p+q)^2][m_{T_Q}^2 - p^2]}
 \Bigg\{G_1(Q^2)(2p+q)_\rho\Bigg[g_{\alpha\beta}-\frac{p_\alpha p_\beta}{m_{T_Q}^2}
 -\frac{(p+q)_\alpha (p+q)_\beta}{m_{T_Q}^2}+\frac{(p+q)_\alpha p_\beta}{2m_{T_Q}^4}\nonumber\\
 & \times (Q^2+2m_{T_Q}^2)
 \Bigg]
 + G_2 (Q^2) \Bigg[q_\alpha g_{\rho\beta}  
 - q_\beta g_{\rho\alpha}-
\frac{p_\beta}{m_{T_Q}^2}  \big(q_\alpha p_\rho - \frac{1}{2}
Q^2 g_{\alpha\rho}\big) 
+
\frac{(p+q)_\alpha}{m_{T_Q}^2}  \big(q_\beta (p+q)_\rho+ \frac{1}{2}
Q^2 g_{\beta\rho}\big) 
\nonumber\\
&-  
\frac{(p+q)_\alpha p_\beta p_\rho}{m_{T_Q}^4} \, Q^2
\Bigg]
-\frac{G_3(Q^2)}{m_{T_Q}^2}(2p+q)_\rho \Bigg[
q_\alpha q_\beta -\frac{p_\alpha q_\beta}{2 m_{T_Q}^2} Q^2 
+\frac{(p+q)_\alpha q_\beta}{2 m_{T_Q}^2} Q^2
-\frac{(p+q)_\alpha q_\beta}{4 m_{T_Q}^4} Q^4\Bigg]
\Bigg\}\,.
\end{align}

Rather than using the previously defined form factors $G_i(Q^2)$, the magnetic and quadrupole form factors, $F_M(Q^2)$ and $F_{\mathcal{D}}(Q^2)$, are experimentally more relevant. These are expressed in terms of $G_i(Q^2)$ as follows:
\begin{align}
\label{edmn07}
&F_M(Q^2) = G_2(Q^2)\,,\nonumber \\
&F_{\cal D}(Q^2) = G_1(Q^2)-G_2(Q^2)+(1+\lambda) G_3(Q^2)\,,
\end{align}
where $\lambda=Q^2/4 m_{T_Q}^2$ is a  kinematic factor  with $Q^2=-q^2$. 
Under standard circumstances, the previous equation is adequate for determining the electromagnetic characteristics of the $T_Q$ states. Nevertheless, because our analysis deals with real photon processes, the form factors need to be evaluated at 
 \( Q^2 = 0 \). In this static limit, the form factors are directly proportional to the familiar static observables: the magnetic ($\mu$) and quadrupole ($\mathcal{D}$) moments. The explicit formulas are:
\begin{align}
\label{edmn08}
&e F_M(0) = 2 m_{T_Q} \mu_{T_Q} \,, \nonumber\\
&e F_{\cal D}(0) = m_{T_Q}^2 {\cal D}_{T_Q}\,.
\end{align}

Accordingly, the formulas for the magnetic and quadrupole moments are obtained, thereby finalizing the hadronic representation. The next stage consists of analyzing the correlation function from the perspective of quark-gluon dynamics and their relevant characteristics.

 \subsection{QCD formulation}

Within the QCD framework for the correlation function, the interpolating currents of the relevant hadronic states are inserted as in Eq. (\ref{edmn01}). Wick’s theorem is then employed to carry out all contractions required to derive the corresponding expressions. The QCD formulation of the correlation functions for these hadronic states is consequently expressed as follows:
%
\begin{align}
\label{eq:QCDSide}
\Pi _{\alpha \beta }^{\mathrm{QCD}-1}(p,q)&=-\frac{i}{2}\int d^{4}xe^{ip\cdot x}   \langle 0 \mid  
\Bigg\{
\mathrm{Tr}\Big[  \gamma _{\alpha } S_{Q}^{aa^{\prime}}(x) \gamma _{\beta }  S_{u}^{b^{\prime }b}(-x)\Big]  
\mathrm{Tr}\Big[ \gamma _{5}{S}_{u}^{bb^{\prime }}(x)\gamma _{5} S_{s}^{b^{\prime }b}(-x)\Big]    
\nonumber\\
&+
\mathrm{Tr}\Big[  \gamma _{\alpha } S_{Q}^{aa^{\prime}}(x) \gamma _{\beta }  S_{d}^{b^{\prime }b}(-x)\Big]  
\mathrm{Tr}\Big[ \gamma _{5}{S}_{d}^{bb^{\prime }}(x)\gamma _{5} S_{s}^{b^{\prime }b}(-x)\Big]  
\Bigg\}\mid 0 \rangle_{F} ,  \\
%
\Pi _{\alpha \beta }^{\mathrm{QCD}- 2}(p,q)&=-\frac{i}{2}\int d^{4}xe^{ip\cdot x}   \langle 0 \mid  
\Bigg\{
\mathrm{Tr}\Big[  \gamma _{5 } S_{Q}^{aa^{\prime}}(x) \gamma _{5}  S_{u}^{b^{\prime }b}(-x)\Big]  
\mathrm{Tr}\Big[ \gamma _{\alpha}{S}_{u}^{bb^{\prime }}(x)\gamma _{\beta} S_{s}^{b^{\prime }b}(-x)\Big]    
\nonumber\\
&+
\mathrm{Tr}\Big[  \gamma _{5 } S_{Q}^{aa^{\prime}}(x) \gamma _{5 }  S_{d}^{b^{\prime }b}(-x)\Big]  
\mathrm{Tr}\Big[ \gamma _{\alpha}{S}_{d}^{bb^{\prime }}(x)\gamma _{\beta} S_{s}^{b^{\prime }b}(-x)\Big]  
\Bigg\}\mid 0 \rangle_{F}. \label{eq:QCDSide1}
 \end{align}
In this context, $\Pi_{\alpha \beta}^{\mathrm{QCD}-1}(p,q)$ is derived for the $D^\ast  K$ and $B^\ast  K$ molecular states, whereas $\Pi_{\alpha \beta}^{\mathrm{QCD}-2}(p,q)$ is obtained for the $DK^\ast  $ and $BK^\ast  $ states.  
For both heavy and light quark fields, the propagator structures take the following form~\cite{Balitsky:1987bk, Belyaev:1985wza}:
\begin{align}
\label{edmn13}
S_{q}(x)&= S_q^{free}(x) 
-\frac {i g_s }{16 \pi^2 x^2} \int_0^1 du \, G^{\mu \nu} (ux)
\bigg[\bar u \rlap/{x} 
\sigma_{\mu \nu} + u \sigma_{\mu \nu} \rlap/{x}
 \bigg],\\
%
S_{Q}(x)&=S_Q^{free}(x)
-i\frac{m_{Q}\,g_{s} }{16\pi ^{2}}  \int_0^1 du \,G^{\mu \nu}(ux)\bigg[ (\sigma _{\mu \nu }{\xslash}
+{\xslash}\sigma _{\mu \nu }) 
    \frac{K_{1}\big( m_{Q}\sqrt{-x^{2}}\big) }{\sqrt{-x^{2}}}
 +2\sigma_{\mu \nu }K_{0}\big( m_{Q}\sqrt{-x^{2}}\big)\bigg],
 \label{edmn14}
\end{align}%
with  
\begin{align}
 S_q^{free}(x)&=\frac{1}{2 \pi x^2}\Big(i \frac{\xslash}{x^2}- \frac{m_q}{2}\Big),\\
 S_Q^{free}(x)&=\frac{m_{Q}^{2}}{4 \pi^{2}} \Bigg[ \frac{K_{1}\big(m_{Q}\sqrt{-x^{2}}\big) }{\sqrt{-x^{2}}}
+i\frac{{\xslash}~K_{2}\big( m_{Q}\sqrt{-x^{2}}\big)}
{(\sqrt{-x^{2}})^{2}}\Bigg],
\end{align}
where $G^{\mu\nu}$ denotes the gluon field-strength tensor, and $K_i$'s represent modified Bessel functions of the second kind.

The contributions in Eqs.~(\ref{eq:QCDSide})–(\ref{eq:QCDSide1}) can be classified into two categories, depending on how the photon engages with the quark lines. The first category represents perturbative interactions via the conventional QED vertex, often called short-distance interactions. The second category involves non-perturbative interactions, described by the photon light-cone distribution amplitudes, corresponding to long-distance effects. 
To achieve a complete and internally consistent theoretical description, both types of contributions must be included. The implementation of these effects is carried out according to the following steps:
\begin{itemize}
 \item For the purpose of accounting for perturbative effects, the subsequent formulation is sufficient.
\begin{align}
\label{free}
S_{Q(q)}^{free}(x) \longrightarrow \int d^4z\, S_{Q(q)}^{free} (x-z)\,\rlap/{\!A}(z)\, S_{Q(q)}^{free} (z)\,.
\end{align}
In this method, one quark propagator—either light or heavy—is inserted into the above expression, while the free part of the other propagators remains unchanged.   

\item The incorporation of non-perturbative contributions in this framework is realized through the following expression:
 \begin{align}
\label{edmn21}
S_{q,\alpha\beta}^{ab}(x) \longrightarrow -\frac{1}{4} \big[\bar{q}^a(x) \Gamma_i q^b(0)\big]\big(\Gamma_i\big)_{\alpha\beta},
\end{align}
where $\Gamma_i = \{\mathbb{1}$, $\gamma_5$, $\gamma_\alpha$, $i\gamma_5 \gamma_\alpha$, $\sigma_{\alpha\beta}/2\}$.  The computational procedure in this framework involves substituting one quark propagator while retaining the full structure of the other propagators. The resulting non-perturbative effects give rise to matrix elements such as $\langle \gamma(q)|\bar{q}(x) \Gamma_i G_{\alpha\beta}q(0)|0\rangle$ and $\langle \gamma(q)|\bar{q}(x) \Gamma_i q(0)|0\rangle$, which are expressed through photon distribution amplitudes (DAs). These DAs capture essential non-perturbative QCD contributions and are necessary for an accurate evaluation of the correlation function. Their explicit forms, including contributions up to twist-4, are detailed in Ref. \cite{Ball:2002ps}. 
In this work, the photon DAs considered involve only the contributions of light quarks. While, in principle, photons could also be emitted at long distances from heavy quarks, such effects are expected to be negligible for the present study. 
Technically, the matrix elements of nonlocal operators are represented through DAs, quark condensates, and certain nonperturbative parameters. Since the influence of these parameters is already small for light quarks, their role becomes even less significant for heavy quarks. In particular, heavy quark condensates scale as \(1/m_c\) and are therefore strongly suppressed~\cite{Antonov:2012ud}. 
For this reason, we omit long-distance contributions associated with heavy quark DAs and retain only the short-distance photon emission from heavy quarks, as given in Eq.~(\ref{free}). 
The methodology for consistently integrating perturbative and non-perturbative contributions is outlined in Refs.~\cite{Ozdem:2022vip,OZDEM:2024jlw, Ozdem:2022eds}. Incorporating these theoretical elements leads to the complete QCD formulation of the correlation function.

\end{itemize}

\subsection{Sum rules for magnetic and quadrupole moments}

By identifying the coefficients of the unique Lorentz structures—$(q_\alpha \varepsilon_\beta - \varepsilon_\alpha q_\beta)$ for the magnetic moment and $(\varepsilon \cdot p) q_\alpha q_\beta$ for the quadrupole moment—in both QCD and hadronic sides, the LCSR are derived. Using these sum rules, one can extract the magnetic and quadrupole moments of the associated molecular tetraquark states in terms of QCD and hadronic parameters, as well as the auxiliary quantities $\mathrm{s_0}$ and $\mathrm{M^2}$. The obtained results are shown below:
\begin{align}
\label{jmu1}
 &\mu^{J_\alpha^1}_{T_Q} \, \lambda^{2}_{{T_Q}}\, e^{-\frac{m_{{T_Q}}^{2}}{\mathrm{M^2}}}   =  \rho_1(\mathrm{M^2},\mathrm{s_0}),~~~~~~~~~~
  \mathcal{D}^{J_\alpha^1}_{T_Q} \, \lambda^{2}_{{T_Q}}\, e^{-\frac{m_{{T_Q}}^{2}}{\mathrm{M^2}}}  =  \rho_2(\mathrm{M^2},\mathrm{s_0}),  \\
 & \mu^{J_\alpha^2}_{T_Q} \, \lambda^{2}_{{T_Q}} \,e^{-\frac{m_{{T_Q}}^{2}}{\mathrm{M^2}}} =  \rho_3(\mathrm{M^2},\mathrm{s_0}), ~~~~~~~~~~
 \mathcal{D}^{J_\alpha^2}_{T_Q} \, \lambda^{2}_{{T_Q}} \,e^{-\frac{m_{{T_Q}}^{2}}{\mathrm{M^2}}}  = \rho_4(\mathrm{M^2},\mathrm{s_0}).\label{jmu4}
 \end{align}
For completeness, the explicit forms of the spectral functions $ \rho_i(\mathrm{M^2},s_0) $ are given in the Appendix.

\end{widetext}

\section{Numerical Evaluations}\label{numerical}

The extraction of magnetic and quadrupole moments through LCSR depends on the specification of various input parameters. The present analysis is carried out using the following set of values:  $m_u =m_d = 0$, $m_s = 93.5 \pm 0.08$ MeV, $m_c = 1.273 \pm 0.0046\,\mbox{GeV}$, $m_b = 4.183 \pm 0.007\,\mbox{GeV}$~\cite{ParticleDataGroup:2024cfk},  
  $m_{D K^\ast  }=2.457^{+0.064}_{-0.068}$ GeV~\cite{Zhou:2025yjb}, 
  $m_{D^\ast   K}=2.538^{+0.059}_{-0.062}$ GeV~\cite{Zhou:2025yjb}, 
  $m_{B K^\ast  }=6.050^{+0.062}_{-0.064}$ GeV~\cite{Zhou:2025yjb}, 
  $m_{B^\ast   K}=6.158^{+0.061}_{-0.063}$ GeV~\cite{Zhou:2025yjb}, 
  $\lambda_{D K^\ast  }=(3.88^{+0.22}_{-0.24}) \times 10^{-3}$ GeV$^5$~\cite{Zhou:2025yjb}, 
  $\lambda_{D^\ast   K}=(4.89^{+0.27}_{-0.29}) \times 10^{-3}$ GeV$^5$~\cite{Zhou:2025yjb},
  $\lambda_{B K^\ast  }=(2.05^{+0.09}_{-0.09}) \times 10^{-2}$ GeV$^5$~\cite{Zhou:2025yjb}, 
  $\lambda_{B^\ast   K}=(2.60^{+0.10}_{-0.10}) \times 10^{-2}$ GeV$^5$~\cite{Zhou:2025yjb},
  $\langle \bar uu\rangle = \langle \bar dd\rangle=(-0.24 \pm 0.01)^3\,\mbox{GeV}^3$, 
  $\langle \bar ss\rangle = (0.8 \pm 0.1)\, \langle \bar uu\rangle$ $\,\mbox{GeV}^3$ \cite{Ioffe:2005ym}, $\chi= -2.85 \pm 0.5 $ GeV$^{-2}$ \cite{Rohrwild:2007yt}, $f_{3\gamma}=-(0.0039 \pm 0.0020)~$GeV$^2$~\cite{Ball:2002ps}, and $\langle g_s^2G^2\rangle = 0.48 \pm 0.14~ \mbox{GeV}^4$~\cite{Narison:2018nbv}. Photon DAs represent one of the main inputs in our numerical calculations. Their explicit forms, as well as other relevant parameters, are presented in Ref.~\cite{Ball:2002ps}.

According to Eqs.~(\ref{jmu1})--(\ref{jmu4}), our calculations critically depend on two auxiliary parameters: $\mathrm{s_0}$ (continuum threshold) and $\mathrm{M^2}$ (Borel mass). For the results to be both reliable and consistent, it is necessary to determine a working region, defined as the parameter range where the computed magnetic and quadrupole moments do not change significantly under small variations. $\mathrm{s_0}$ is best understood as a physical threshold scale, beyond which continuum and excited state contributions to the correlation function play a dominant role. Although different prescriptions for fixing its working region can be found in the literature, phenomenological evidence consistently favors the interval $ (m_{H} + 0.4)^2~\mathrm{GeV}^2 \leq \mathrm{s_0} \leq (m_{H} + 0.8)^2~\mathrm{GeV}^2$, where $m_H$ denotes the mass of the hadronic state under consideration.  We rely on this interval in our calculations because it has been shown to yield robust results in analogous theoretical approaches. Two fundamental constraints govern the choice of the working region for $\mathrm{M^2}$: the pole dominance (PC) criterion, which requires the ground-state term to outweigh the continuum and excited-state contributions, and the convergence of the OPE (CVG), which ensures the reliability of the truncated expansion. These constraints can be formulated as:
\begin{align}
 \mbox{PC}  &=\frac{\rho_i (\mathrm{M^2},\mathrm{s_0})}{\rho_i (\mathrm{M^2},\infty)} > 35 \%, ~~~~~~
%
 \mbox{CVG} =\frac{\rho_i^{\mathrm{Dim 7}} (\mathrm{M^2},\mathrm{s_0})}{\rho_i (\mathrm{M^2},\mathrm{s_0})} < 5\%,
 \end{align}
 where $ \rho_i^{\text{Dim7}}( \mathrm{M^2}, \mathrm{s_0})$ represents the term with the highest dimension in the QCD formulation of $ \rho_i ( \mathrm{M^2}, \mathrm{s_0}) $.
 
\begin{figure}[htb!]
\includegraphics[width=0.4\textwidth]{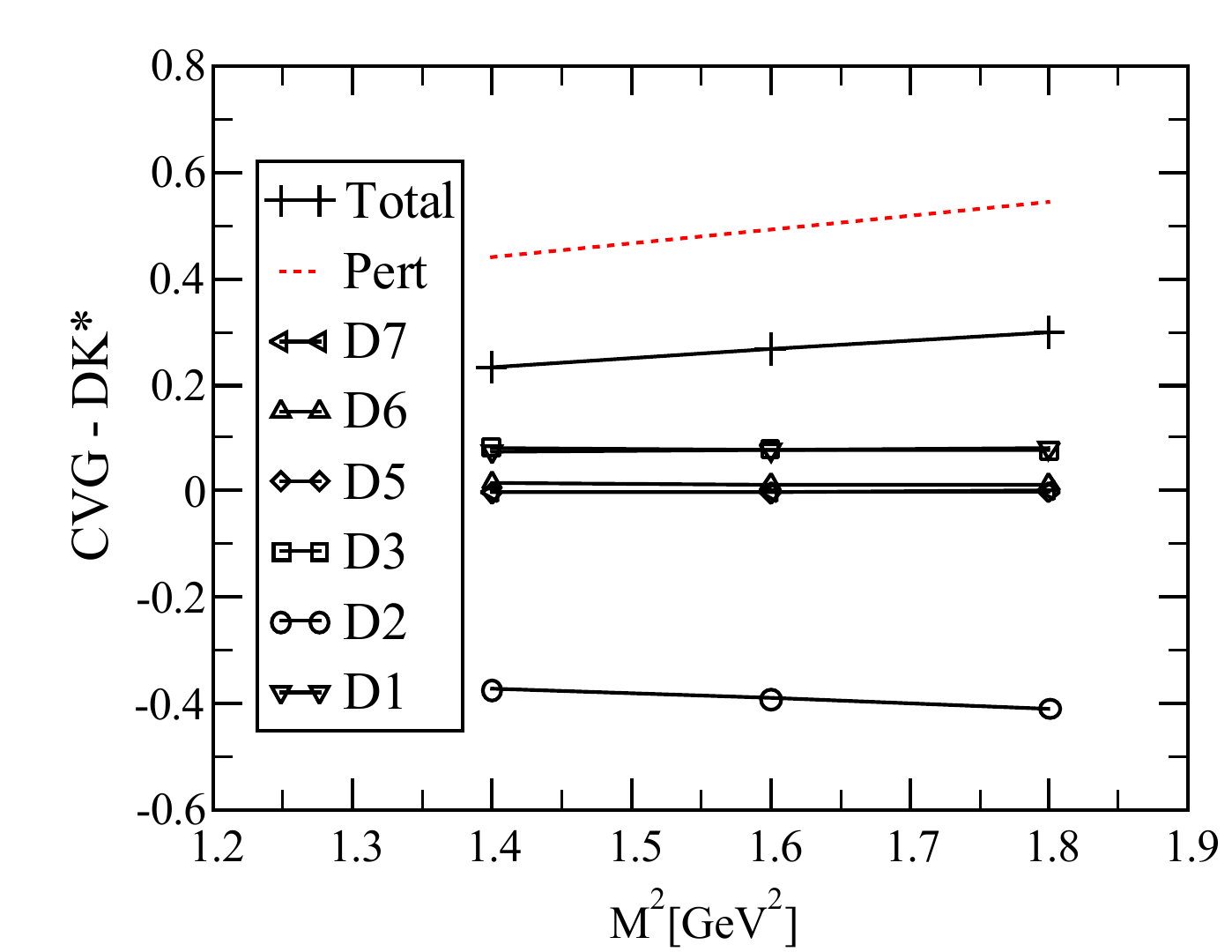}~~~~
\includegraphics[width=0.4\textwidth]{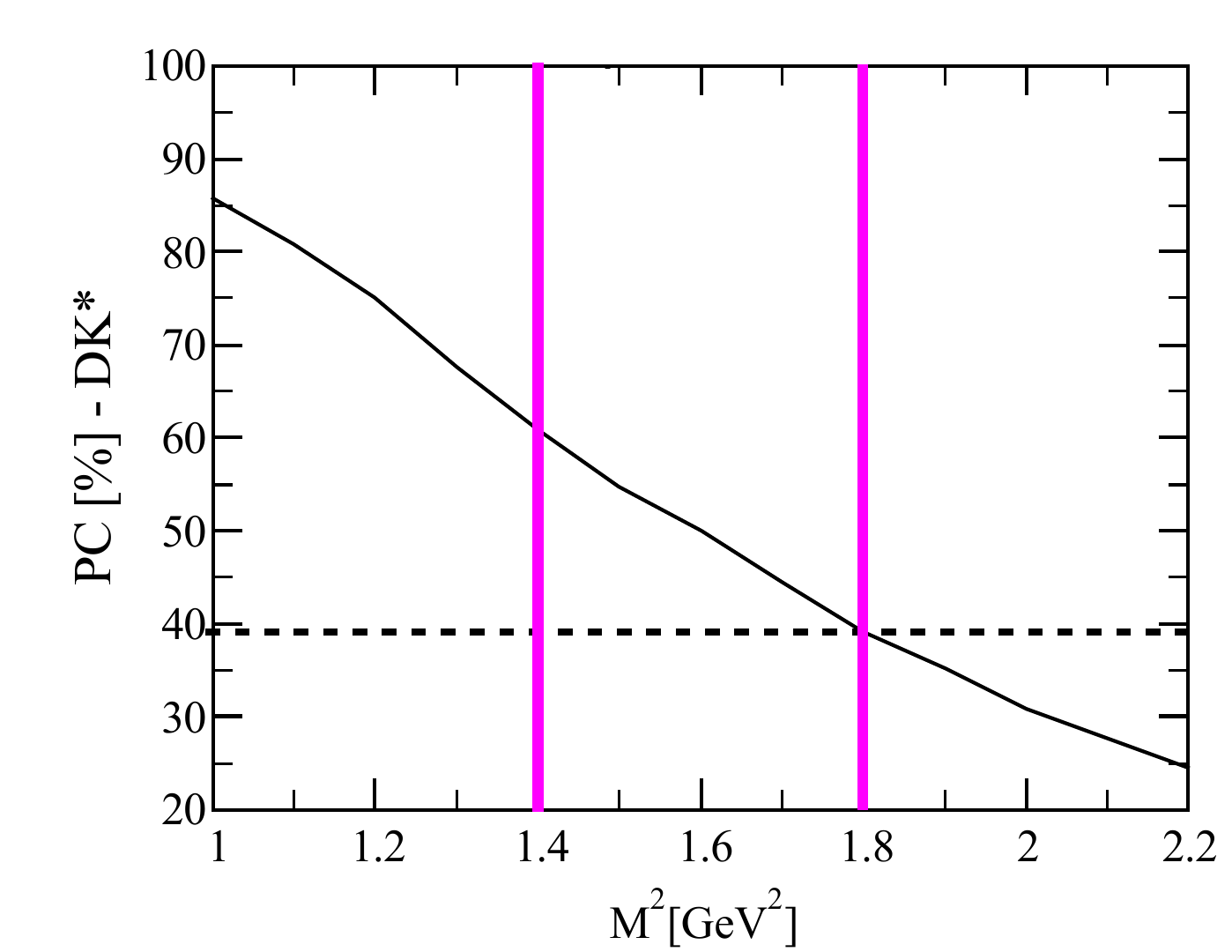}\\
\caption{CVG analysis (upper panel) and PC analysis (lower panel) for the magnetic dipole moment of the $D K^\ast$ state as a function of $\mathrm{M^2}$ at fixed $\mathrm{s_0}$ values. In the lower panel, the vertical lines indicate the adopted Borel window, whereas the horizontal line represents the smallest PC value extracted within this region in the present study.}
 \label{Msqfig1}
  \end{figure}

\begin{widetext}
 
     \begin{figure}[htp]
\centering
  \includegraphics[width=0.4\textwidth]{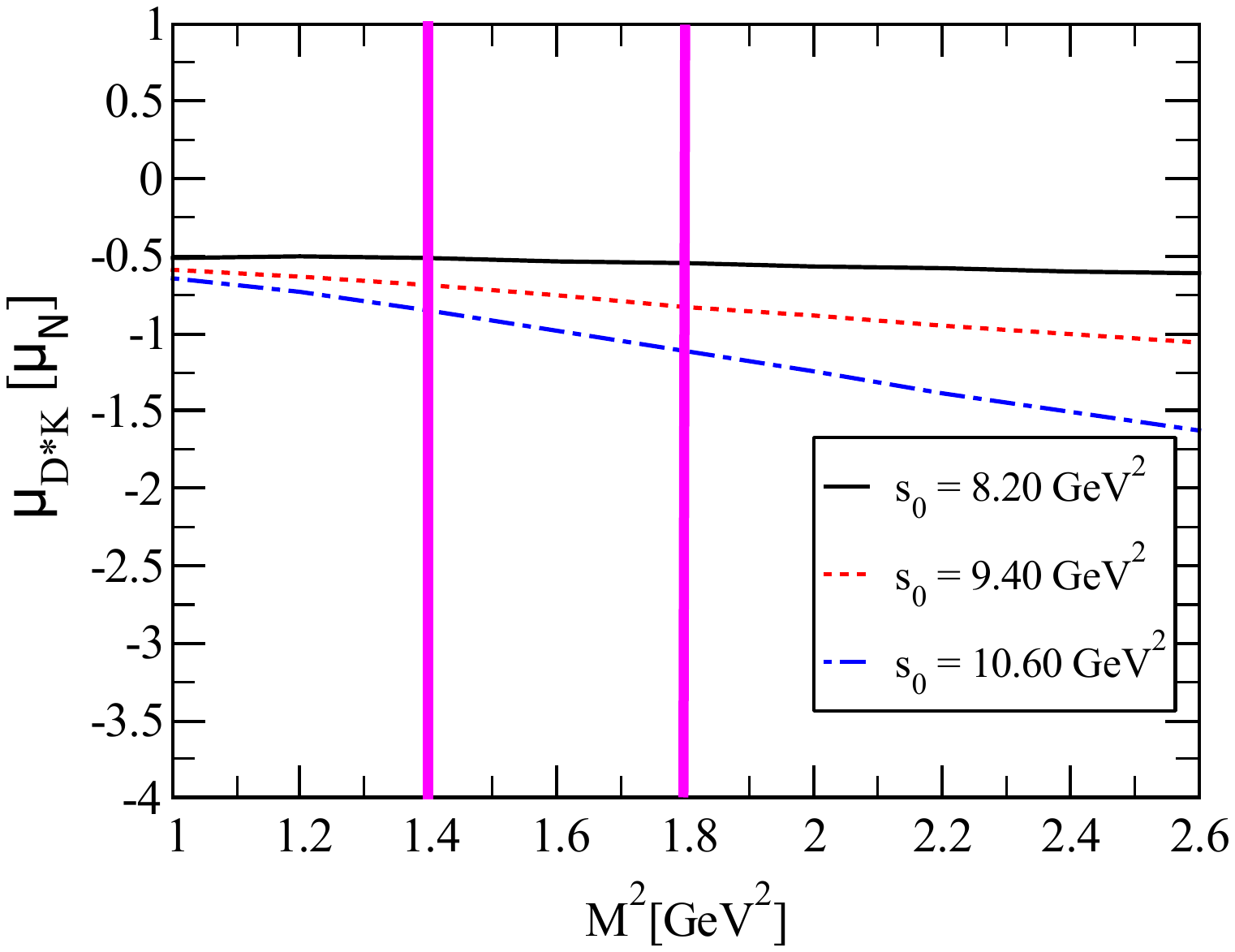} ~~~~~
  \includegraphics[width=0.4\textwidth]{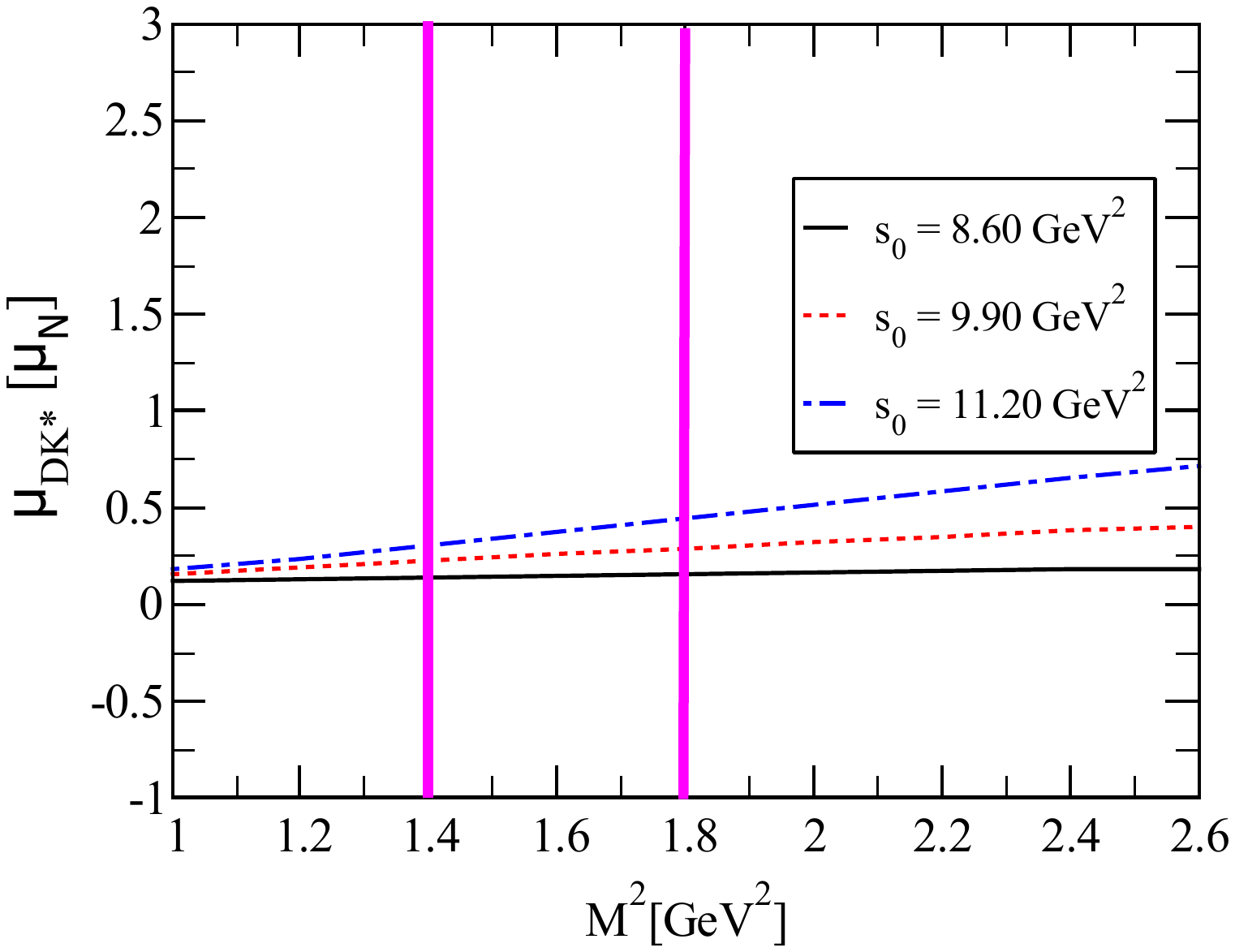} \\
    \includegraphics[width=0.4\textwidth]{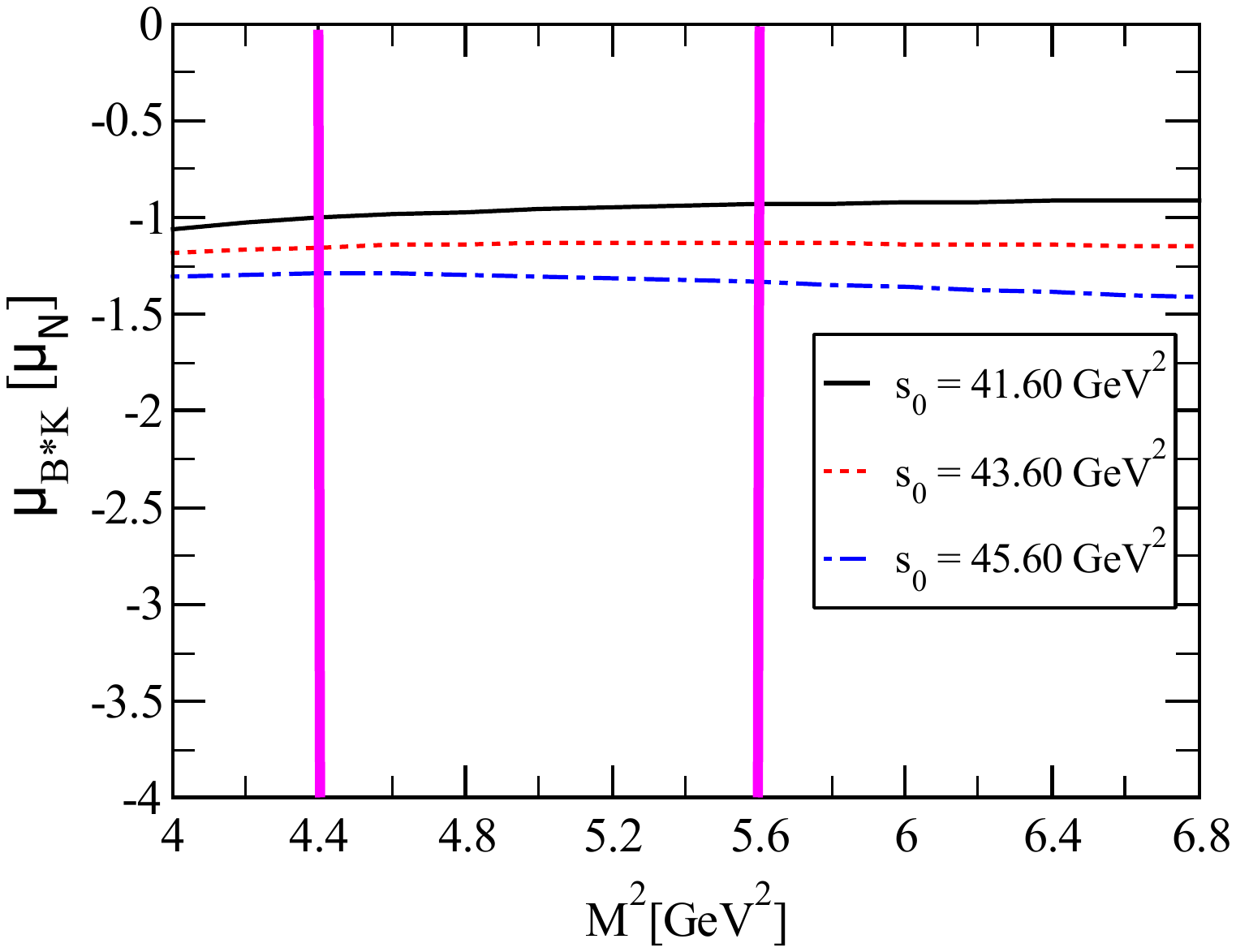}~~~~~
    \includegraphics[width=0.4\textwidth]{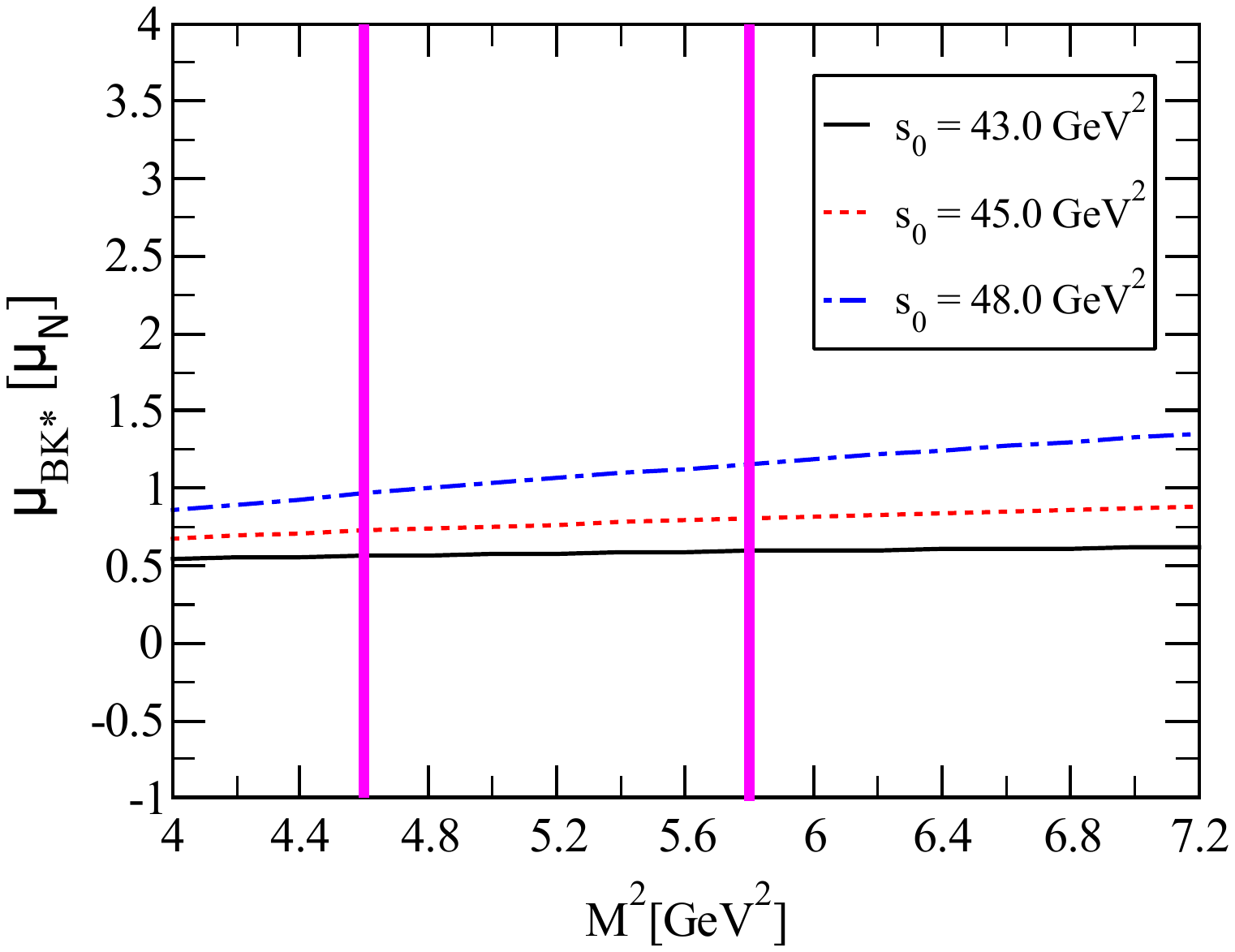}
  \caption{Magnetic moments of the molecular $D^\ast K$, $D K^\ast $, $B^\ast K$ and  $BK^\ast $ states as a function of $\mathrm{M^2}$ at several continuum thresholds. The area bounded by the vertical lines corresponds to the adopted Borel window.}
 \label{figMsq}
  \end{figure}
  
  \end{widetext}

 Based on the prescribed conditions, the appropriate domains for the auxiliary parameters are identified and reported in Table~\ref{table}.  As shown in Table ~\ref{table}, the Borel windows and threshold parameters are chosen to simultaneously guarantee OPE convergence and the dominance of the pole contribution over the continuum. 
 For all states, the PC falls within the 40--65\% range in the working Borel window. These significant PC values demonstrate that the ground-state pole dominates the correlation function, validating the single-pole approximation employed in the hadronic representation and confirming that our interpolating currents effectively isolate the lowest-lying state from the continuum. 
 Furthermore, the convergence of the operator product expansion, quantified as the ratio of higher-dimensional contributions to the total, remains below 0.25\% for all cases. This excellent convergence ensures the truncation of the OPE series at dimension-7 is justified and that non-perturbative contributions are under control.  
The simultaneous satisfaction of these two criteria---dominant pole contribution and solid OPE convergence---within the same Borel window establishes a self-consistent framework for reliably extracting the electromagnetic properties of these molecular states.  
 As an illustration, for the $D K^\ast$ state, Figure~\ref{Msqfig1} demonstrates that within the adopted Borel window, the pole contribution significantly exceeds the continuum contribution, thereby confirming the dominance of the ground state. Moreover, the relative magnitudes of the condensate terms indicate a satisfactory convergence of the OPE series. 
 For completeness, Fig.~\ref{figMsq} displays how the extracted magnetic moments vary with changes in the auxiliary parameters. As expected, the results show only a mild sensitivity within the chosen ranges. Nevertheless, a degree of uncertainty remains, stemming from residual parameter dependencies.

 By carrying out a detailed numerical analysis, we have determined the magnetic and quadrupole moments of the molecular states $D^\ast   K$, $D K^\ast$, $B^\ast K$, and $B K^\ast  $. The results, summarized in Table~\ref{table}, incorporate full error propagation from all input parameters, with systematic treatment of the variations in auxiliary parameters. The estimated contributions to the total uncertainty are distributed approximately as follows: 13\% from the tetraquark masses, 22\% from their residues, 30\% from the continuum threshold $\mathrm{s_0}$, 7\% from the Borel mass parameter $\mathrm{M^2}$, 11\% from the photon DAs, and the remaining 17\% from other sources.  
  
\begin{widetext}

  \begin{table}[htb!]
	\addtolength{\tabcolsep}{6pt}
	\caption{Working regions of $\mathrm{s_0}$ and $\mathrm{M^2}$ together with the PC and CVG for the electromagnetic properties of the molecular $D^\ast K$, $D K^\ast$, $B^\ast K$ and  $BK^\ast $ states.}
	\label{table}
	\begin{ruledtabular}
\begin{tabular}{lccccccc}
	   \\
States & $\mathrm{s_0}\,\,[\mathrm{GeV}^2]$&   $\mathrm{M^2}\,\,[\mathrm{GeV}^2]$& PC\,\,[$\%$] & CVG\,\,[$\%$]  & $\mu\, [\mu_N]$ &  $\mathcal{D}~[\times 10^{-2}\, \mbox{fm}^{2}]$  \\
	   \\
	   \hline\hline
	  \\
$D^\ast   K$& [8.20, 10.6]&  [1.4, 1.8]& [68.46, 39.11]& $ 0.10$ & $- 1. 01 \pm 0.40$ & $-0.19 \pm 0.05$ 
\\
\\
$B^\ast   K$& [41.6, 45.6]&  [4.4, 5.6]& [64.87, 42.67]& $ 0.11$ & $- 1.14 \pm 0.25$ & $-0.10 \pm 0.02$ \\
\\
$D K^\ast  $& [8.60, 11.2]&  [1.4, 1.8]& [60.78, 39.13]& $0.22$ & $~~0.28 \pm 0.12$ & $~~0.51 \pm 0.11$\\
\\
$B K^\ast  $& [43.0, 48.0]&  [4.6, 5.8]& [64.06, 41.98]& $0.20$ & $~~0.86 \pm 0.30$ & $~~0.39 \pm 0.05$\\
\\
\end{tabular}
\end{ruledtabular}
\end{table}

 \end{widetext}

\begin{widetext}
 
     \begin{figure}[htp]
\centering
  \includegraphics[width=0.85\textwidth]{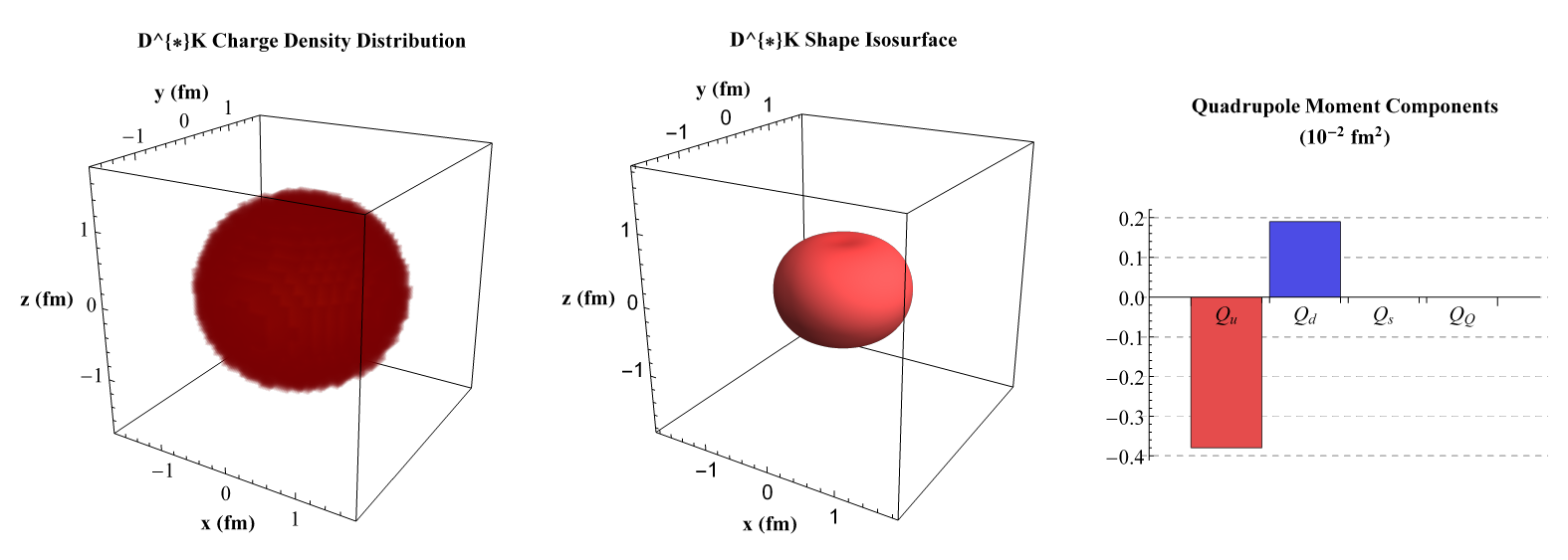} \\
  \includegraphics[width=0.85\textwidth]{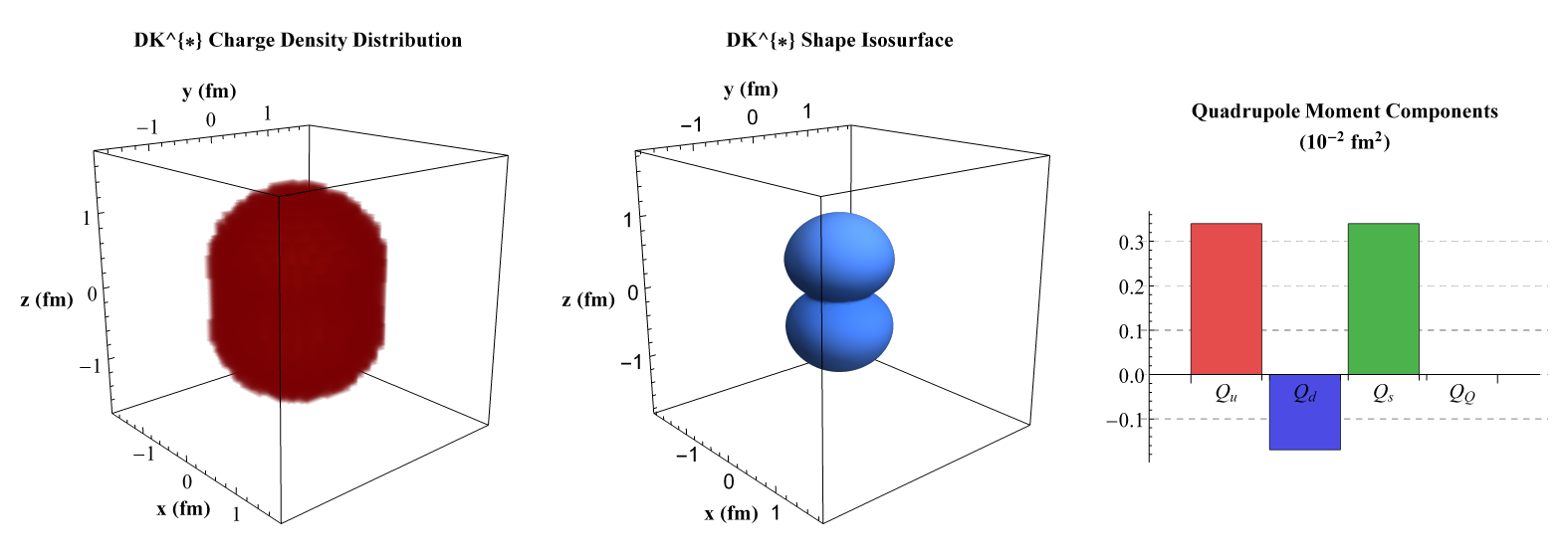} \\
    \includegraphics[width=0.85\textwidth]{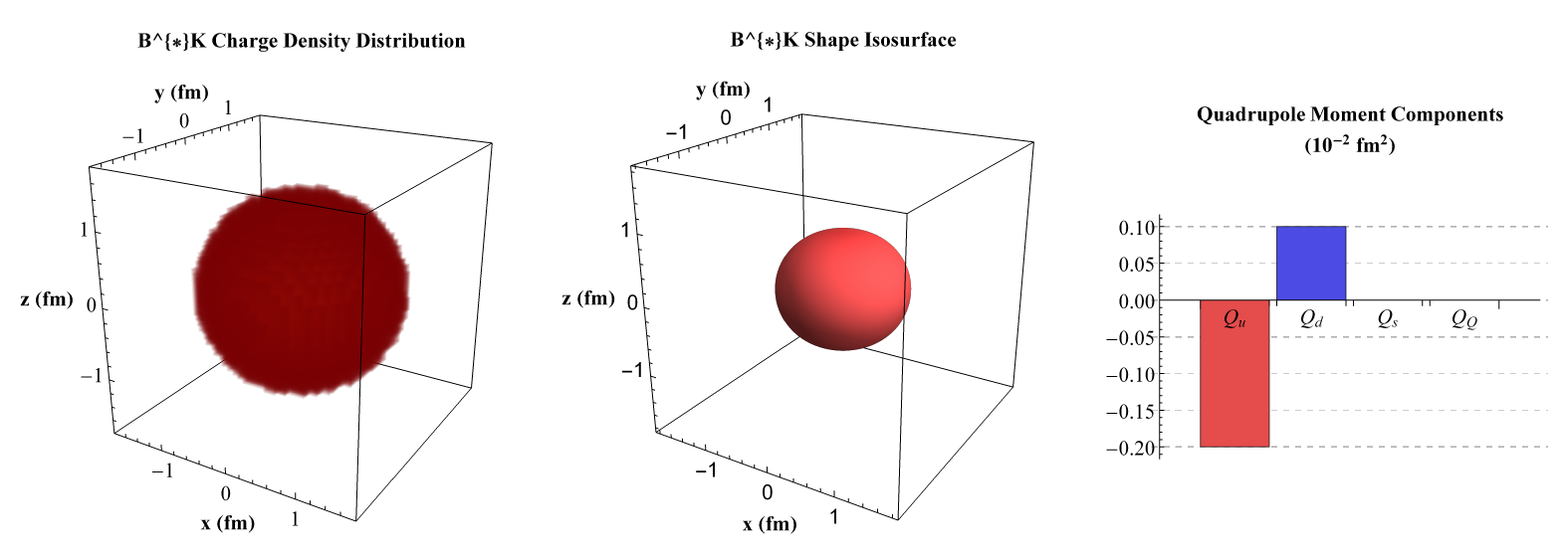}\\
    \includegraphics[width=0.85\textwidth]{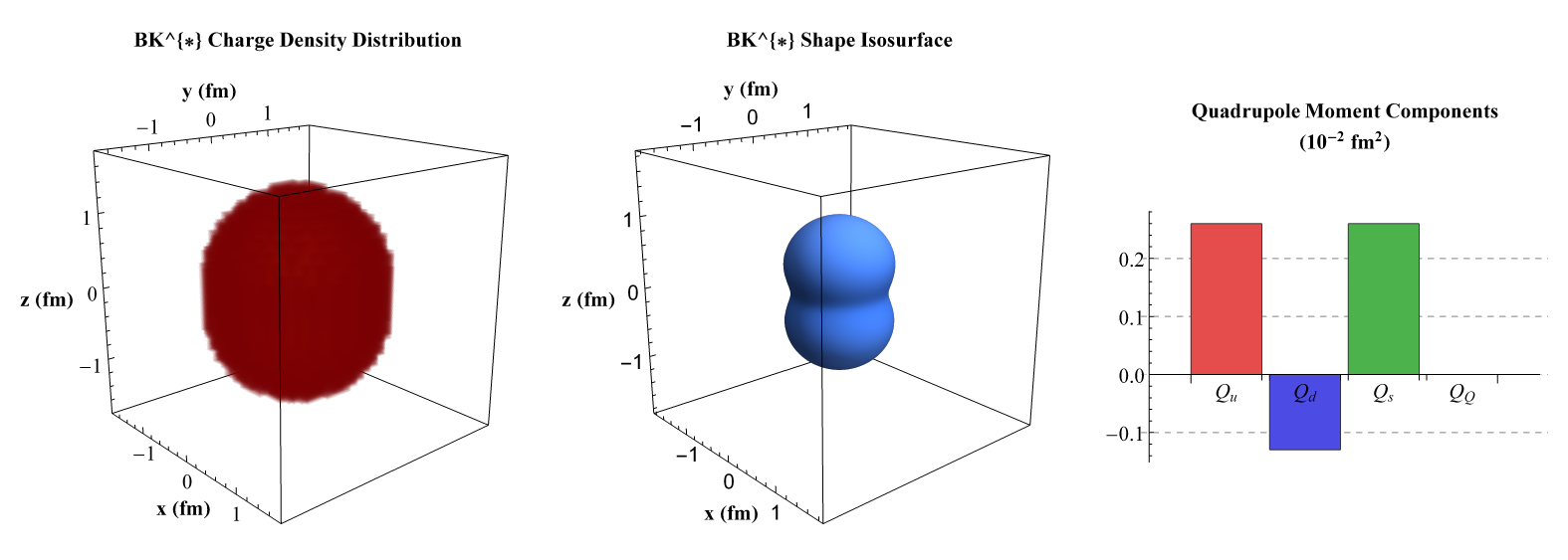}
\caption{Three-dimensional visualizations of $D^\ast K$, $D K^\ast$, $B^\ast K$, and $BK^\ast$ states. 
Left: charge density $\rho$ (e/fm$^3$); 
Middle: isosurface at 5\% of the peak charge density for each state; 
Right: quadrupole moment components (fm$^2$). 
These plots illustrate quadrupole-induced deformations, providing a clear view of the spatial geometry and charge distribution. All axes are in femtometers (fm).}
 \label{figMsq1}
  \end{figure}
  
  \end{widetext}

Our analysis leads to the following main observations:
   
\begin{itemize}

\item The extracted magnetic dipole moments of the $D^\ast  K$, $DK^\ast  $, $B^\ast  K$, and $BK^\ast  $ molecular states display a clear pattern reflecting their internal spin structure and flavor composition (see Tables~\ref{table}–\ref{table3}). The $D^\ast  K$ and $B^\ast  K$ states carry negative magnetic moments, $\mu(D^\ast  K) = -1.01 \pm 0.40\,\mu_N$ and $\mu(B^\ast  K) = -1.14 \pm 0.25\,\mu_N$. Their flavor decomposition reveals that the large negative $u$-quark contribution dominates, partially cancelled by a positive $d$-quark term, with $\mu_u/\mu_d \approx -2.0$ consistent with the expected isospin structure. The heavy-quark contribution is small but nonzero (negative for charm, positive for bottom), while the strange quark remains inactive due to the pseudoscalar kaon. In contrast, the $DK^\ast  $ and $BK^\ast  $ states exhibit positive moments, $\mu(DK^\ast  ) = 0.28 \pm 0.12\,\mu_N$ and $\mu(BK^\ast  ) = 0.86 \pm 0.30\,\mu_N$. In $DK^\ast  $, the moment is mainly driven by the $u$ quark with a smaller positive $s$-quark contribution ($\mu_u/\mu_s \approx 3.6$) and partial cancellation from the $d$ quark. For $BK^\ast  $, the $u$ and $s$ quarks contribute comparably ($\mu_u/\mu_s \approx 1.25$), jointly producing the positive moment. Heavy-quark contributions in these states are negligible.  
This $K$–$K^\ast  $ dichotomy illustrates a key organizing principle: in $K$-containing states, the pseudoscalar kaon acts as a spin-inactive spectator and the magnetic moment is governed by the vector meson’s light and heavy quarks, while in $K^\ast  $-containing states, the vector kaon actively drives the electromagnetic response through its constituent $u$ and $s$ quarks, with the pseudoscalar meson ($D$, $B$) playing the role of a spectator.

The quadrupole moments provide crucial insight into the spatial charge distribution of the molecular states. The slightly negative values for the $D^*K$ and $B^*K$ states suggest a mild oblate (flattened) deformation, while the positive values for the $DK^*$ and $BK^*$ states indicate a prolate (elongated) shape. These deformations are dominated by the light and strange quarks, with negligible contributions from heavy quarks. 
Although the magnitudes are small---on the order of $10^{-2}~\text{fm}^2$---the consistent signs across states confirm that the oblate/prolate distinction is a robust prediction of our analysis. It should be noted, however, that the orientation of these deformations relative to the spin axis is model-dependent and not directly determined here. 
As illustrated in Fig.~\ref{figMsq1}, the charge density and isosurface visualizations make these deformation patterns visually clear, reinforcing the interpretation of the quadrupole moments. 
Experimentally, these shape differences could manifest in polarization observables---such as photon angular distributions in radiative decays or spin asymmetries in electroproduction. Facilities like BESIII, LHCb, GlueX, or future electron-ion colliders may have the precision to detect these subtle effects, especially when combining differential cross-section and polarization data.

\item The flavor decomposition presented in Table~\ref{table3} reveals a definitive hierarchy in the electromagnetic response: for the $K$-containing states ($D^*K$, $B^*K$), the $u$-quark provides the dominant negative contribution, which is partially canceled by a positive $d$-quark term, while the $s$-quark in the pseudoscalar kaon remains inactive. In contrast, for the $K^*$-containing states ($DK^*$, $BK^*$), both the $u$- and $s$-quarks within the vector meson provide significant positive contributions. It is crucial to emphasize that these values represent the effective contribution of each quark flavor, obtained by isolating terms proportional to its electric charge ($e_u$, $e_d$, $e_s$) in the sum rules while setting others to zero. Consequently, they incorporate the quark's spatial distribution and interactions within the composite state and should not be equated with bare or constituent quark magnetic moments.

\item The electromagnetic features carry important experimental implications. Large positive magnetic moments of $DK^\ast  $ and $BK^\ast  $ suggest stronger interactions with external electromagnetic fields, enhancing visibility in photo- and electro-production experiments. The sign reversal between $D^\ast  K$/$B^\ast  K$ and $DK^\ast  $/$BK^\ast  $ moments provides a clear experimental signature for distinguishing internal structures. Quadrupole deformations could manifest in polarization observables, offering additional handles to test the molecular structure.

\item The distinct magnetic dipole moments of the $D^\ast  K$, $DK^\ast  $, $B^\ast  K$, and $BK^\ast  $ molecular states suggest clear experimental signatures. The negative moments of $D^\ast  K$ and $B^\ast  K$, dominated by the $u$ quark with partial cancellation from the $d$ quark, indicate that measurements of photoproduction or radiative transitions should reflect this destructive interference among light-quark contributions. In contrast, the positive moments of $DK^\ast  $ and $BK^\ast  $ arise from the constructive interplay of $u$ and $s$ quarks in the vector $K^\ast  $, predicting enhanced transition rates and cross sections in channels involving these states. Lattice QCD could test these predictions directly by probing the flavor decomposition, such as the isospin ratio $\mu_u/\mu_d \simeq -2$ in $D^\ast  K$ or the dominant $u$-quark contribution in $DK^\ast $ and the more balanced $u$- and $s$-quark contributions in $BK^\ast $. Overall, these patterns provide a robust framework for experimentally discriminating molecular configurations from compact tetraquark scenarios, in which the flavor hierarchy and magnetic moments would be expected to differ significantly.

\item To the best of our knowledge, no prior theoretical estimates exist for these electromagnetic moments. Comparisons with compact tetraquark models suggest key differences: in a diquark--antidiquark picture, quark contributions would be more symmetric, sign patterns could differ, and quadrupole moments would be closer to zero. Observing the predicted sign reversal of magnetic moments along with significant quadrupole deformations would strongly favor the molecular interpretation over a compact tetraquark structure. Future lattice QCD or alternative current studies could further test these structural predictions.

\end{itemize}

\begin{widetext}

  \begin{table}[htb!]
	\addtolength{\tabcolsep}{6pt}
	\caption{Flavor-decomposed magnetic and quadrupole ($\times 10^{-2}$) moments contributions for the $D^\ast  K$, $D K^\ast $, $B^\ast  K$ and  $BK^\ast $  molecular states.}
	\label{table3}
	\begin{ruledtabular}
\begin{tabular}{lccccccc}
	   \\
States &   $\mu_{u} \,[\mu_N]$& $\mu_d \,[\mu_N]$ & $\mu_s \,[\mu_N]$& $\mu_{Q} \,[\mu_N]$&$\mu_{tot} \,[\mu_N]$\\
	   \\
	   \hline\hline
	  \\
$D^\ast  K$&  $-1.54$& $~~0.77$& $0.00$&$-0.24$ &$-1.01$\\
\\
$B^\ast  K$&  $-2.52$& $~~1.26$& $0.00$ & $~~0.12$ &$-1.14$\\
\\
$DK^\ast $&  $~~0.36$& $-0.18$& $0.10$ & $~~0.00$&~~0.28\\
\\
$BK^\ast $&  $~~0.66$& $-0.33$& $ 0.53$ & $~~ 0.00$&~~0.86\\
\\
\hline\hline 
\\
States &   $\mathcal D_{u} \,[\mathrm{fm}^2]$& $\mathcal D_d \,[\mathrm{fm}^2]$&    $\mathcal D_s \,[\mathrm{fm}^2]$ & $\mathcal D_{Q} \,[\mathrm{fm}^2]$&$\mathcal D_{tot} \,[\mathrm{fm}^2]$\\
	   \\
	   \hline\hline
	  \\
$D^\ast  K$&  $-0.38$& $~~0.19$& $0.00$ &$0.00$ &$-0.19$
\\
\\
$B^\ast  K$&  $-0.20$& $~~0.10$& $0.00$&$0.00$ &$-0.10$\\
\\
$DK^\ast $&  $~~0.34$& $-0.17$& $0.34$ &$0.00$&$~~0.51$\\
\\
$BK^\ast $&  $~~0.26$& $-0.13$& $0.26$ &$0.00$&$~~0.39$\\
\\
\end{tabular}
\end{ruledtabular}
\end{table}

\end{widetext}

\subsection*{Experimental Implications and Observables} \label{subsec:experimental}

The magnetic and quadrupole moments calculated here are not merely static properties but directly influence dynamic experimental observables. In photoproduction reactions (e.g., $\gamma p \to D^*K X$ or $\gamma p \to BK^* X$ at facilities like GlueX or future electron-ion colliders), the cross-section near threshold is sensitive to the electromagnetic structure of the produced state. A larger magnetic moment $|\mu|$ typically enhances the photoproduction cross-section due to stronger coupling to the photon field. More specifically, the sign of the magnetic moment can manifest in polarization observables. For instance, the beam asymmetry or the polarization transfer from a circularly polarized photon beam to the produced state can depend on the relative sign between the charge and magnetic couplings. The clear sign reversal we predict between $D^*K$/$B^*K$ (negative $\mu$) and $DK^*$/$BK^*$ (positive $\mu$) should therefore lead to measurably different polarization patterns, providing a distinctive experimental signature to discriminate between these molecular configurations. 
Furthermore, the magnetic moment directly governs the rate of radiative transitions between spin-aligned and spin-anti-aligned states. For a magnetic dipole (M1) transition between members of the same spin-multiplet, the partial width $\Gamma_{M1}$ is proportional to $|\mu|^2 \cdot E_\gamma^3$, where $E_\gamma$ is the photon energy. Our predictions can thus be tested by measuring radiative decay widths, for example, in the decays of excited bottom-strange states observed by LHCb. The significant difference in $|\mu|$ between $B^*K$ ($\sim 1.14 \mu_N$) and $BK^*$ ($\sim 0.86 \mu_N$) should translate into a factor of $\sim 1.7$ difference in their respective M1 transition rates, a clear testable prediction.

The quadrupole moment $\mathcal{D}$, while challenging to measure directly, influences the angular distribution of decay products. In the photoproduction of a spin-1 state with a non-zero $\mathcal{D}$, the alignment of the spin axis relative to the beam direction is modified, which in turn affects the angular distribution of its decay products (e.g., $D^*K \to D_s \pi$ or $BK^* \to B_s \pi$). Measurements of these angular distributions in electroproduction experiments, where the virtual photon's polarization can be controlled, could provide indirect access to the quadrupole deformation.

If confirmed experimentally, these patterns would provide strong evidence for the molecular picture, as the predicted magnetic moment signatures and flavor hierarchies are natural consequences of an extended molecular structure but may differ significantly in compact tetraquark scenarios.
 
\section{Discussion and Outlook} \label{sum}

In this work, we have carried out the first systematic calculation of the magnetic and quadrupole moments for the axial-vector molecular candidates $D^\ast K$, $DK^\ast $, $B^\ast K$, and $BK^\ast $, which naturally emerge as the open-charm and open-bottom counterparts of the $D_{s1}(2460)$ and $D_{s1}(2536)$. Our results, obtained within the framework of QCD light-cone sum rules, provide deep insights into the internal electromagnetic structure of these exotic hadrons and offer clear, testable predictions to guide future experimental and theoretical efforts.

The most striking finding is the pronounced dichotomy between states containing a pseudoscalar ($K$) versus a vector ($K^\ast $) meson. The negative magnetic moments of the $D^\ast K$ and $B^\ast K$ systems are unequivocally linked to the antialignment of light-quark spins, dominated by the $u$-quark. In stark contrast, the positive and sizable moments of the $DK^\ast $ and $BK^\ast $ states are driven by the constructive interplay of the $u$ and $s$ quarks within the spin-carrying $K^\ast $ meson. This pattern, complemented by the oblate versus prolate charge distributions revealed by the quadrupole moments, establishes a powerful organizational principle: the electromagnetic identity of a molecular state is dictated by the spin-parity of its constituent mesons. The consistent suppression of heavy-quark contributions across all systems further validates the molecular picture and is in harmony with expectations from heavy-quark symmetry. Beyond their intrinsic theoretical value, these calculated moments serve as crucial benchmarks for distinguishing the molecular interpretation from other structural models, such as compact diquark-antidiquark tetraquarks. The specific flavor hierarchies ($\mu_u/\mu_d \approx -2.0$, comparable $u$- and $s$-quark contributions), sign sequences, and non-zero deformations we predict are natural consequences of a loosely bound, extended molecular configuration but are highly unlikely to emerge from a compact, localized wavefunction.

The electromagnetic moments we obtain provide a quantitative benchmark for testing the molecular structure against other interpretations, notably the compact tetraquark model. In a compact diquark-antidiquark picture, the wave function is typically more symmetric and localized. This would likely lead to significantly different outcomes: the magnetic moments could exhibit altered sign sequences and smaller magnitudes (closer to zero) due to the more confined charge distribution and different spin-flavor correlations. The sizable values we find—e.g., $\mu_{D^{*}K} \sim -1~\mu_N$ and $\mu_{BK^{*}} \sim +0.9~\mu_N$—along with their non-zero quadrupole deformations, are natural consequences of the extended molecular geometry and the coherent sum of the constituent meson moments. These features are challenging to reproduce with a compact, point-like tetraquark configuration.

Our quantitative predictions thus translate a structural hypothesis into a set of experimentally verifiable criteria. The substantial magnitudes of these magnetic moments suggest significant coupling to real and virtual photons, which should lead to observable production rates in high-precision experiments such as BESIII or LHCb, reflecting the sensitivity of production mechanisms to the underlying spin structures of the molecular states.  Observation of production yields consistent with these predictions would provide supporting evidence for the molecular interpretation and the predicted flavor decomposition, whereas significant deviations could indicate the presence of alternative structural configurations, such as compact tetraquark arrangements. 
On the theoretical front, the most definitive validation of our predictions must come from lattice QCD simulations. We strongly encourage the lattice community to compute these electromagnetic properties directly. This could be achieved by employing the external field method or by calculating the $F_M(Q^2)$ and $F_{\mathcal{D}}(Q^2)$ form factors at low $Q^2$ from three-point functions, followed by an extrapolation to $Q^2 = 0$. Such first-principles calculations would not only test our specific results but also provide a model-independent decomposition of the light ($u, d, s$) and heavy ($c, b$) quark contributions. This would offer the most stringent test of the underlying flavor dynamics and the molecular hypothesis presented in this work.

In conclusion, the electromagnetic properties calculated herein are not merely static numbers but dynamic fingerprints that encode the inner workings of these exotic states. Future confirmations of these predictions through photo-/electro-production experiments, radiative transition measurements, and lattice QCD computations will be instrumental in moving beyond the identification of new hadronic states and toward a definitive understanding of their nature. This work provides a concrete framework that bridges theoretical construction with experimental discovery.

\begin{widetext}

\appendix 

\section*{Appendix: Explicit expressions for $\rho_i(\mathrm{M^2},\mathrm{s_0})$  functions }
The obtained sum rules for the magnetic  and quadrupole moments of $D^\ast   K$, $B^\ast   K$, $DK^\ast  $ and $BK^\ast  $ states for are presented
as follows:
\begin{align}
 \rho_1(\mathrm{M^2},\mathrm{s_0})&= -\frac{m_Q^2}{8192 \pi^6} \Big[2(e_u+e_d )
  \big(7 m_Q^{10} I[-2] + 27 m_Q^6 I[0] - 2 m_Q^4 I[1] + 32 I[3]\big) + 
 e_Q \big(m_Q^{10} I[-2] + 32 m_Q^8 I[-1] \nonumber\\
    &- 6 m_Q^6 I[0] + 
    40 m_Q^4 I[1] - 3 m_Q^2 I[2] + 64 I[3]\big)\Big]
    \nonumber\\
    &+ \frac{    \langle g_s^2 G^2\rangle \langle \bar q q \rangle }{  73728 m_Q \pi^4}  (e_u+e_d ) \Big[ -12 m_Q^2 \mathbb A[u_ 0] I[
   0] + (13 I_ 4[\mathcal S] + 
    12 (-I_ 4[\mathcal {\tilde S}] + 2
       I_ 6[\varphi_ {\gamma}])) (2 m_Q^2 I[0] - I[1]) 
       \Big]
    \nonumber\\
    & +\frac{   \langle g_s^2 G^2\rangle f_{3 \gamma} }{294912 m_Q^2 \pi^4} (e_u+e_d ) \Big[ -13  I_ 1[\mathcal V] (2 m_Q^6 I[-1] + m_Q^4 I[0] + I[2]) + 
 24  (2 m_Q^8 I[-2] + m_Q^4 I[0]  \nonumber\\
    &- 4 m_Q^2 I[1]+ 
    I[2])I_ 5[\psi^a] + 48 (2 m_Q^8 I[-2] + 3 m_Q^4 I[0] - 2 m_Q^2 I[1] + 
     I[2]) \psi^a[u_ 0] 
    \Big]
    \nonumber\\
 %
    &+ \frac{    \langle g_s^2 G^2\rangle \langle \bar q q \rangle  \chi}{  3072 m_Q \pi^4}  (e_u+e_d ) \Big[  
  (2 m_Q^6 I[-1] + m_Q^4 I[0] + I[2]) \varphi_ {\gamma}[u_ 0] \Big]
    \nonumber\\
    &+\frac{ m_Q \langle \bar q q \rangle }{1024 \pi^4} (e_u+e_d ) \Big[ 6 m_Q^4 I_ 6 (m_Q^4 I[-2] - 2 m_Q^2 I[-1] + I[0])[h_ {\gamma}] + 
 6  (m_Q^6 IS[-1] - m_Q^4 I[0])\mathbb A[u_ 0] - 
 3 m_Q^2 (I_ 4[\mathcal S] \nonumber\\
    &- 
    I_ 4[\mathcal {\tilde S}]) (m_Q^4 I[-1] - 2 m_Q^2 I[0] + I[1]) \Big]
    \nonumber\\
    &+\frac{ m_Q \langle \bar q q \rangle  \chi }{256 \pi^4} (e_u+e_d ) \Big[  
  (m_Q^8 I[-1] + 2 m_Q^6 I[0] + m_Q^4 I[1] + 
     4 I[3]) \varphi_ {\gamma}[u_ 0] \Big]
      \nonumber\\
    &+\frac{f_{3\gamma}}{2048 \pi^4} (e_u+e_d )  \Big[    (m_Q^{10} I[-2] + 2 m_Q^8 I[-1] + m_Q^6 IS[0] + 
    4 I[3]) I_ 5[\psi^a] + 
 I_ 1[\mathcal V] (m_Q^8 I[-1] + 2 m_Q^6 I[0] \nonumber\\
    &+ m_Q^4 I[1] + 
    4 I[3]) + 4 m_Q^6 (m_Q^4 IS[-2] - I[0]) \psi^a[u_ 0] \Big], \\
 \rho_2(\mathrm{M^2},\mathrm{s_0})&=  \frac{  m_Q  \langle g_s^2 G^2\rangle \langle \bar q q \rangle }{ 73728 \pi^4}  (e_u+e_d ) \Big[m_Q^2 (I_4[\mathcal T_2] + I_4[\mathcal T_4]) I[-2] - (26 I_4[\mathcal T_1] + 
    27 I_4[\mathcal T_2] - 24 I_4[\mathcal T_3] - 23 I_4[\mathcal T_4]) I[-1] \Big]
    \nonumber\\
    & +\frac{   \langle g_s^2 G^2\rangle f_{3 \gamma} }{73728  m_Q^2 \pi^4}   \Big[ 
    (13(e_u -2e_s+ e_d) I_ 4[\mathcal A] +
   96 (e_d + e_u)  I_ 6[\psi_ {\gamma}^{\nu}] ) (2 m_Q^2 I[0] - I[1])
    \Big]
    \nonumber\\
    &-\frac{ 3m_Q \langle \bar q q \rangle }{256 \pi^4} (e_u+e_d ) \Big[  (I_ 4[\mathcal T_ 1] + I_ 4[\mathcal T_ 2] - I_ 4[\mathcal T_ 3] - 
   I_ 4[\mathcal T_ 4]) (m_Q^6 I[-2] + 2 m_Q^4 I[-1] + m_Q^2 I[0] + 
   4 I[1]) \Big]
      \nonumber\\
    &-\frac{3 m_Q^4 f_{3\gamma}}{1024 \pi^4} (e_u+e_d )  \Big[  8 m_Q^2  (m_Q^4 I[-3] - 2 m_Q^2 I[-2] + 
    I[-1]) I_ 6[\psi_ {\gamma}^{\nu}] +  (m_Q^4 I[-2] - 2 m_Q^2 I[-1] + I[0])I_ 4[\mathcal A] \Big], \\
%
 \rho_3
 (\mathrm{M^2},\mathrm{s_0})&= \frac{m_Q^2}{4096 \pi^6} (e_u-2e_s +e_d )
 \Big[ 7 m_Q^{10} I[-2] + 27 m_Q^6 I[0] - 2 m_Q^4 I[1] + 32 I[3] \Big]
    \nonumber\\
    &+ \frac{m_s \langle g_s^2 G^2\rangle \langle \bar q q \rangle }{  6144 m_Q^2 \pi^4} (e_u +e_d ) \Big[  -m_Q^2 \mathbb A[u_ 0] I[0] + 2
 I_ 6[h_ {\gamma}] (-2 m_Q^2 I[0] +  I[1])  \Big]
    \nonumber\\
    & +\frac{   \langle g_s^2 G^2\rangle f_{3 \gamma} }{294912  m_Q^2 \pi^4}(e_u - 2 e_s + e_d)\Big[  
 13(2 m_Q^6 I[-1] + m_Q^4 I[0]  
      + I[2])I_ 1[\mathcal V]  + 
 48  (2 m_Q^8 I[-2] - m_Q^4 I[0] \nonumber\\
    &- I[2]) \psi^
    a[u_ 0]
    \Big]
    \nonumber\\
    &+ \frac{m_s \langle g_s^2 G^2\rangle \langle \bar q q \rangle \chi}{  3172 m_Q^2 \pi^4} (e_u +e_d ) \Big[   (2 m_Q^6 I[-1] + m_Q^4 I[0] + I[2]) \varphi_ {\gamma}[u_ 0]  \Big]
    \nonumber\\
    &+\frac{ m_s \langle \bar q q \rangle }{512 m_Q^2 \pi^4} (e_u +e_d ) \Big[ 6 m_Q^8  (m_Q^2 I[-2] - I[-1])I_ 6[h_ {\gamma}] +  (3 m_Q^8 I[-1] - m_Q^6 I[0] + 3 m_Q^4 I[1] - 6 m_Q^2 I[2] \nonumber\\
    & + 
    I[3])\mathbb A[
   u_ 0] ) \varphi_ {\gamma}[u_ 0] \Big]\nonumber \\
%
    &-\frac{ m_s \langle \bar q q \rangle \chi }{128 m_Q^2 \pi^4} (e_u +e_d ) \Big[ 
m_Q^2 (m_Q^8 I[-1] + m_Q^6 I[0] + 
     2 I[3]) \varphi_ {\gamma}[u_ 0] \Big]
      \nonumber\\
          &-\frac{f_{3\gamma}}{1024 \pi^4}  (e_u - 2 e_s + e_d)\Big[  (m_Q^8 I[-1] +
     m_Q^6 I[0] + 2 I[3])  I_ 1[\mathcal V]  + 
 2  m_Q^6 (m_Q^4 I[-2] - I[0]) \psi^a[u_ 0] \Big],  
      \end{align}
 \begin{align}
    \rho_4(\mathrm{M^2},\mathrm{s_0})&= - \frac{  m_Q  \langle g_s^2 G^2\rangle \langle \bar q q \rangle }{ 73728 \pi^4}  (e_u+e_d ) \Big[(I_4[\mathcal T_1] + I_4[\mathcal T_2]) (m_Q^2 I[-2] - 3 I[-1]) \Big]
    \nonumber\\
    & +\frac{   \langle g_s^2 G^2\rangle f_{3 \gamma} }{36864 m_Q^2 \pi^4} (e_u - 2 e_s + e_d)\Big[ 
    13 (2 m_Q^2 I[0] - I[1]) I_4[\mathcal A]
    \Big]
      \nonumber\\
    &+\frac{3 m_Q^6 f_{3\gamma}}{512 \pi^4} (e_u-2e_s+e_d )  \Big[    (m_Q^2 I[-2] - I[-1])I_ 4[\mathcal A] + 
 4   (m_Q^4 I[-3]  - 
    2 m_Q^2 I[-2] + I[-1])I_ 6[\psi_ {\gamma}^{\nu}] \Big],
 \end{align}

\noindent where $\langle g_s^2 G^2 \rangle$ represents the gluon condensate, and $\langle \bar{q} q \rangle$ corresponds to the light-quark condensate.
 The functions $I[n]$, $I_1[\mathcal{A}]$, $I_2[\mathcal{A}]$, $I_3[\mathcal{A}]$, $I_4[\mathcal{A}]$, $I_5[\mathcal{A}]$ and $I_6[\mathcal{A}]$    are defined as:
\begin{align}
I[n]&= \int_{(m_Q+m_s)^2}^{s_0} ds \,s^n e^{-s/\mathrm{M^2}}\nonumber\\
 I_1[\mathcal{A}]&=\int D_{\alpha_i} \int_0^1 dv~ \mathcal{A}(\alpha_{\bar q},\alpha_q,\alpha_g)
 \delta'(\alpha_ q +\bar v \alpha_g-u_0),\nonumber\\
   I_2[\mathcal{A}]&=\int D_{\alpha_i} \int_0^1 dv~ \mathcal{A}(\alpha_{\bar q},\alpha_q,\alpha_g)
 \delta'(\alpha_{\bar q}+ v \alpha_g-u_0),\nonumber\\
   I_3[\mathcal{A}]&=\int_0^1 du~ A(u)\delta'(u-u_0),\nonumber\\
  I_4[\mathcal{A}]&=\int D_{\alpha_i} \int_0^1 dv~ \mathcal{A}(\alpha_{\bar q},\alpha_q,\alpha_g)
 \delta(\alpha_ q +\bar v \alpha_g-u_0),\nonumber\\
   I_5[\mathcal{A}]&=\int D_{\alpha_i} \int_0^1 dv~ \mathcal{A}(\alpha_{\bar q},\alpha_q,\alpha_g)
 \delta(\alpha_{\bar q}+ v \alpha_g-u_0),\nonumber\\
 I_6[\mathcal{A}]&=\int_0^1 du~ A(u),\nonumber
 \end{align}
where $\mathcal{A}$ stands for the corresponding photon DAs. Here, ${\cal D} \alpha_i$ can be written as:
\begin{eqnarray}
\label{nolabel05}
\int {\cal D} \alpha_i = \int_0^1 d \alpha_{\bar q} \int_0^1 d
\alpha_q \int_0^1 d \alpha_g \,  \delta(1-\alpha_{\bar
q}-\alpha_q-\alpha_g)~.
\end{eqnarray}
It is worth noting that the Borel transformations are applied according to the standard expressions  
\begin{align}
 \mathcal{B}\Big\{ \frac{1}{[p^2-m_i^2][(p+q)^2-m_f^2]} \Big\} &\rightarrow e^{-m_i^2/\mathrm{M_1^2} - m_f^2/\mathrm{\mathrm{M_2^2}}}
\end{align}  
in the hadronic representation, and  
\begin{align}
 \mathcal{B}\Big\{ \frac{1}{[m^2 - \bar u\, p^2 - u (p+q)^2]^{\alpha}} \Big\} &\rightarrow (\mathrm{M^2})^{2-\alpha} \delta(u-u_0) e^{-m^2/\mathrm{M^2}}
\end{align}  
in the QCD representation, where  
\begin{align*}
 \mathrm{M^2} = \frac{\mathrm{M_1^2} \mathrm{M_2^2}}{\mathrm{M_1^2}+\mathrm{M_2^2}}, \qquad u_0 = \frac{\mathrm{M_1^2}}{\mathrm{M_1^2}+\mathrm{M_2^2}}.
\end{align*}  
Here, $\mathrm{M_1^2}$ and $\mathrm{M_2^2}$ denote the Borel parameters associated with the initial and final states, respectively. Since the same hadronic state appears in both channels, it is reasonable to adopt $\mathrm{M_1^2} = \mathrm{M_2^2} = 2 \mathrm{M^2}$ and $u_0 = 1/2$. This choice provides a symmetric treatment of both sides of the correlation function and effectively suppresses contributions from higher resonances and the continuum.

\end{widetext}
\bibliography{DstarKmolecule.bib}
\bibliographystyle{elsarticle-num}

\end{document}